\documentclass[12pt,a4paper,dvips,final]{article}
\usepackage{epsfig,times,wrapfig}

\renewcommand{\section}[1]{\noindent\mbox{\bf{#1}} \newline\noindent}
\newcommand{\secref}[1]{\noindent\mbox{\bf{#1}}}
\renewcommand{\subsection}[1]{\noindent\mbox{\it{#1}} \newline\noindent}
\renewcommand{\subsubsection}[1]{\noindent\mbox{#1}--}

\newfont{\tf}{cmbx10 at 16pt}
\newfont{\af}{cmr10 at 14pt}

\def\fwb{75mm}
\def\fhb{62mm}
\def\fwc{73mm}
\def\fhc{60mm}

\newcommand{\gray}{$\gamma$-ray\ }
\begin{document}

\frenchspacing

{\tf \noindent Deciphering diffuse Galactic continuum gamma rays \\}

{\af \noindent I.V. Moskalenko$^{1,2}$ and A.W. Strong$^1$ \\}

{\it \noindent
$^1$Max-Planck-Institut f\"ur extraterrestrische Physik, 
D-85740 Garching, Germany \\
$^2$Institute for Nuclear Physics, Moscow State University, 
119 899 Moscow, Russia }\\

\section{Abstract}
Inverse Compton scattering appears to play a more important role in the
diffuse Galactic continuum emission than previously thought, from  MeV
to GeV energies. We compare models having a large inverse Compton
component with EGRET data, and find good agreement in the longitude and
latitude distributions at low and high energies. We test an alternative
explanation for the $\ge$1 GeV \gray excess, the hard nucleon
spectrum,  using  secondary antiprotons and positrons.
\\

\section{Introduction.}
We are developing a model which aims to reproduce self-consistently
observational data of many kinds related to cosmic-ray origin and
propagation: direct measurements of nuclei, electrons and positrons,
gamma rays, and synchrotron radiation (Strong 1998)\footnote{For more
details see \it
http://www.gamma.mpe--garching.mpg.de/$\sim$aws/aws.html}.

Here we concentrate on the inverse Compton (IC) contribution to the
diffuse Galactic continuum gamma-ray emission.  Recent results from
both COMPTEL and EGRET indicate that IC scattering is a more important
contributor to the diffuse emission that previously believed.  COMPTEL
results (Strong 1997) for the 1--30 MeV range show a latitude
distribution in the inner Galaxy which is broader than that of HI and
H$_2$, so that bremsstrahlung of electrons on the gas does not appear
adequate and a more extended component such as IC is required.  The
broad distribution is the result of the large $z$-extent of the
interstellar radiation field which can interact cosmic-ray electrons up
to several kpc from the plane.  At much higher energies, the puzzling
excess in the EGRET data above 1 GeV relative to that expected for
$\pi^0$-decay has been suggested to orginate in IC scattering (e.g.,
Pohl 1998) from a hard interstellar electron spectrum.  We test  this
scenario with comparisons of the predicted gamma-ray sky with EGRET
data. We also test an alternate hypothesis, the hard nucleons spectrum,
using antiprotons and positrons. 
\\

\section{Models.} 
We consider a propagation model with reacceleration using parameters
derived from isotopic composition (Strong 1998).  Energy losses for
electrons by ionization, Coulomb scattering, bremsstrahlung, IC and
synchrotron are included.  A new calculation of the interstellar
radiation field (ISRF) has been made based on stellar population models
and IRAS and COBE data. An investigation of the effect of the
anisotropy of the ISRF has shown that this has a significant influence
on the intensity and distribution of the IC radiation.  Photons moving
away from the observer are scattered anisotropically, enhancing the
radiation for example at high latitude. This effect is included in our
models.  The $\pi^0$-decay gamma rays are calculated explicitly from
the propagated proton and Helium spectra (Dermer 1986, Moskalenko
1998a).  The electron injection spectral index is taken as --1.8 in the
case of reacceleration models; this value is necessary to obtain
consistency with radio synchrotron spectrum towards the Galactic pole.
Without reacceleration an injection index near --2.0 is required.
Figure~1 shows the electron spectrum at  $R_\odot = 8.5$ kpc in the
disk for these models, and the synchrotron spectrum towards the
Galactic pole.  Following Pohl (1998), for the present study we do not
require consistency with the locally measured electron spectrum above
10 GeV since the rapid energy losses mean that this is not necessarily
representative of even the local interstellar average.  (Agreement with
the locally measured electron spectrum would require a break in the
injection spectrum at a few GeV, as has often been adopted in the
past).  A halo size (distance from plane to boundary) of $z_h$=4 kpc is
adopted, consistent with the $^{10}Be$ analysis in a the accompanying
paper (Strong 1998).
\\

\begin{figure}[t!]
   \begin{picture}(148,69)(5,-59)
      \put(0,0){ \makebox(75,0)[tl]{ \psfig{file=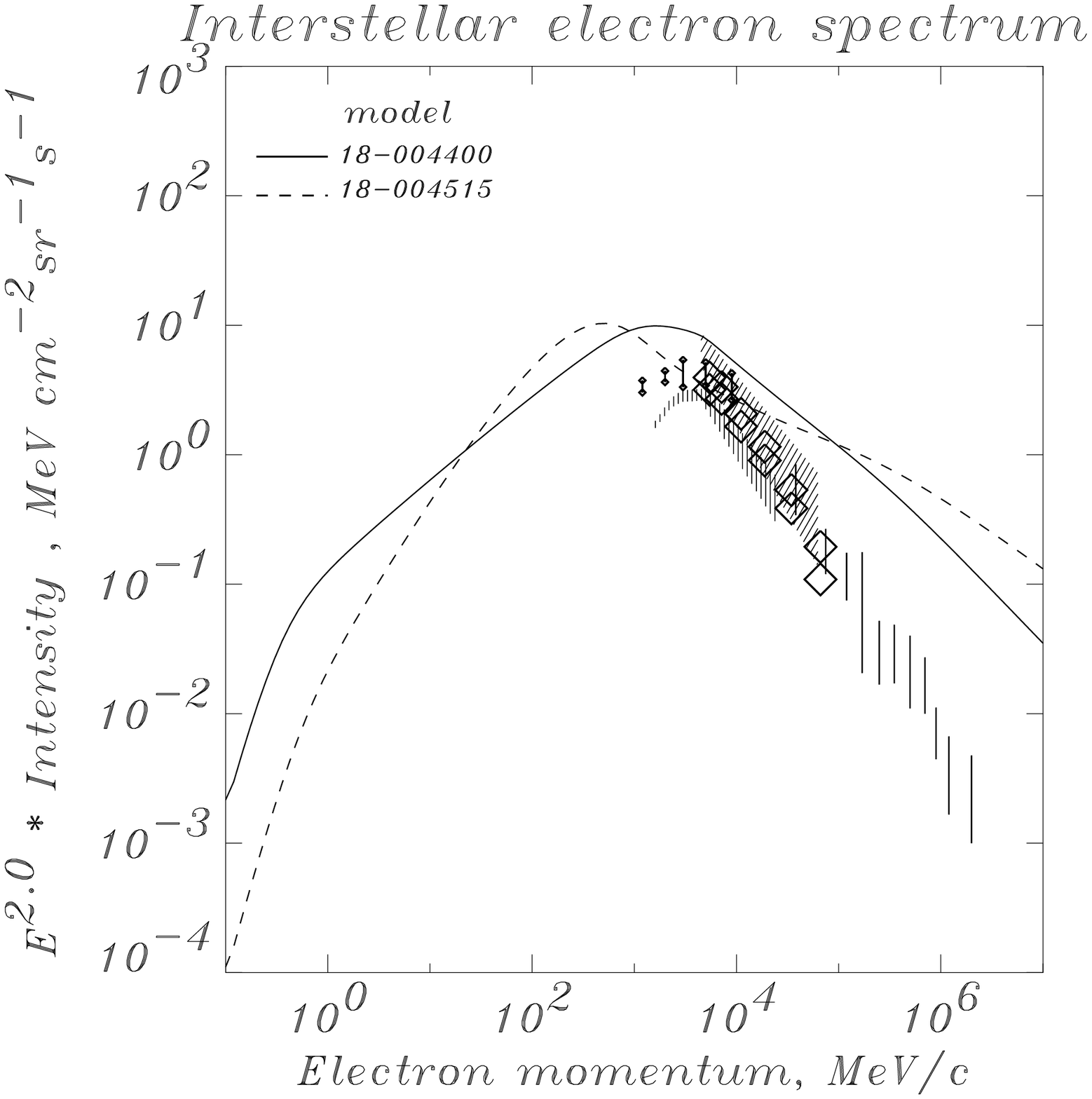,%
         height=\fhb,width=\fwb,clip=}}}
      \put(75,0){ \makebox(75,0)[tl]{ \psfig{file=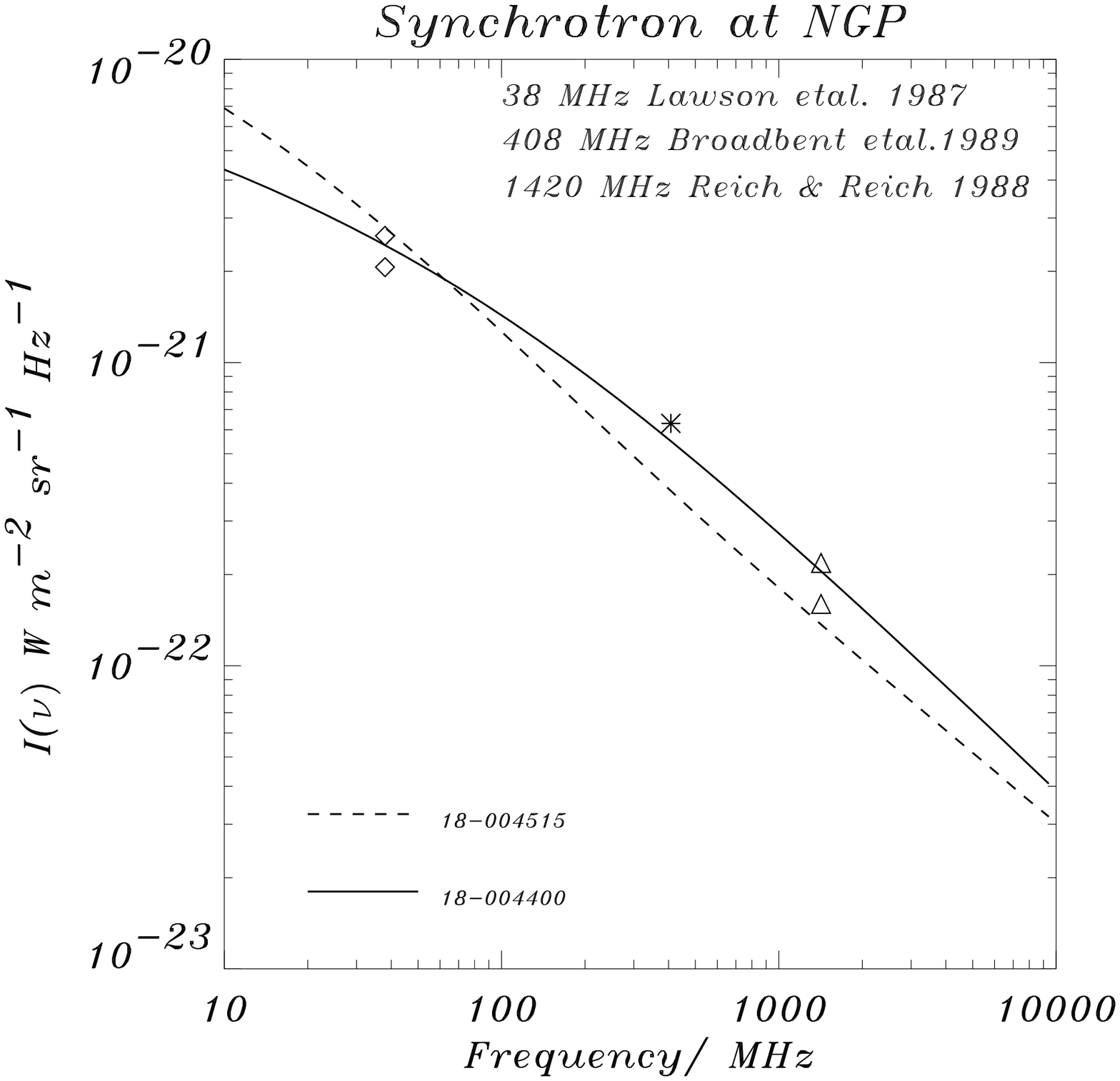,%
         height=\fhb,width=\fwb,clip=}}}
   \end{picture}
\small Fig. 1.
{\it Left:} Electron spectrum at $R_\odot$=8.5 kpc in the plane,
for models with and without reacceleration. Data points:
direct measurements, see references in Moskalenko (1998a).
{\it Right:} Synchrotron spectrum towards NGP for these electron
spectra, compared to observational data.
\end{figure}

\section{Comparison with EGRET data.}
Figure~2 shows the model latitude and longitude \gray distributions
for the inner Galaxy for 70--100 MeV, convolved with the EGRET
point-spread function, compared to EGRET Phase 1--4 data.  The separate
components are also shown.  In this model the contributions from IC,
bremsstrahlung and $\pi^0$-decay are about equal at 100 MeV.  (Note
that point sources such as the Vela pulsar have not been removed from
the data, but we are here only interested in the large-scale profiles).
The  comparison shows that a model with large IC component can
indeed reproduce the data. This energy range is close to that in which
COMPTEL data led to similar conclusions (Strong 1997).  Turning to high
energies, Figure~3 shows profiles for 4000--10000 MeV; again the
comparison shows that the adoption of a hard electron injection spectrum
is a viable explanation for the $>$1 GeV excess. The latitude
distribution here is not as wide as at low energies owing to the rapid
energy losses of the electrons, so that an observational distinction
between a gas-related $\pi^0$-component from a hard nucleon spectrum
and the present IC model does not seem possible.
\\

\begin{figure}[t!]
   \begin{picture}(148,73)(5,-61)
      \put(0,0){ \makebox(75,0)[tl]{ \psfig{file=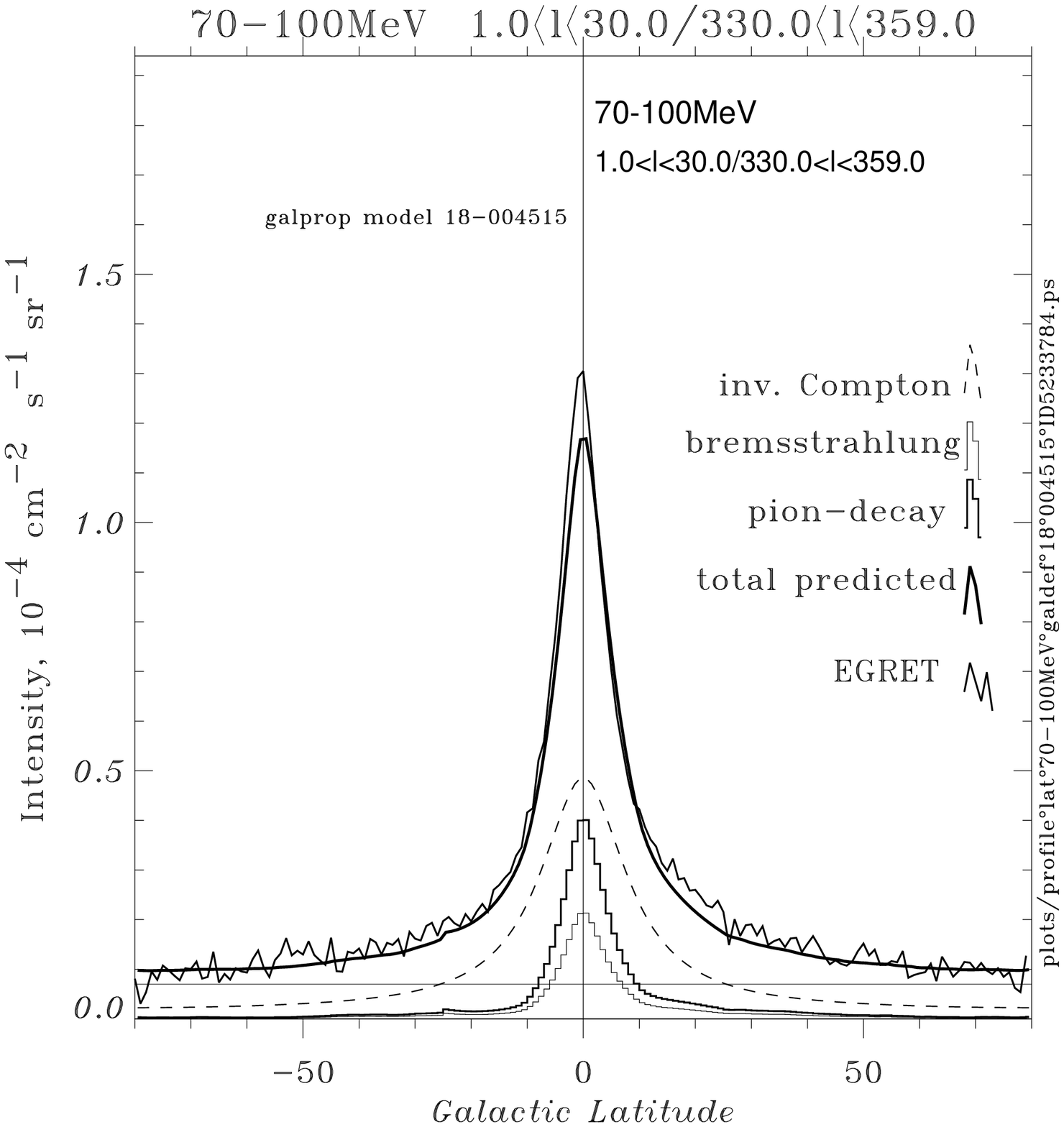,%
         height=\fhb,width=\fwb,clip=}}}
      \put(75,0){ \makebox(75,0)[tl]{ \psfig{file=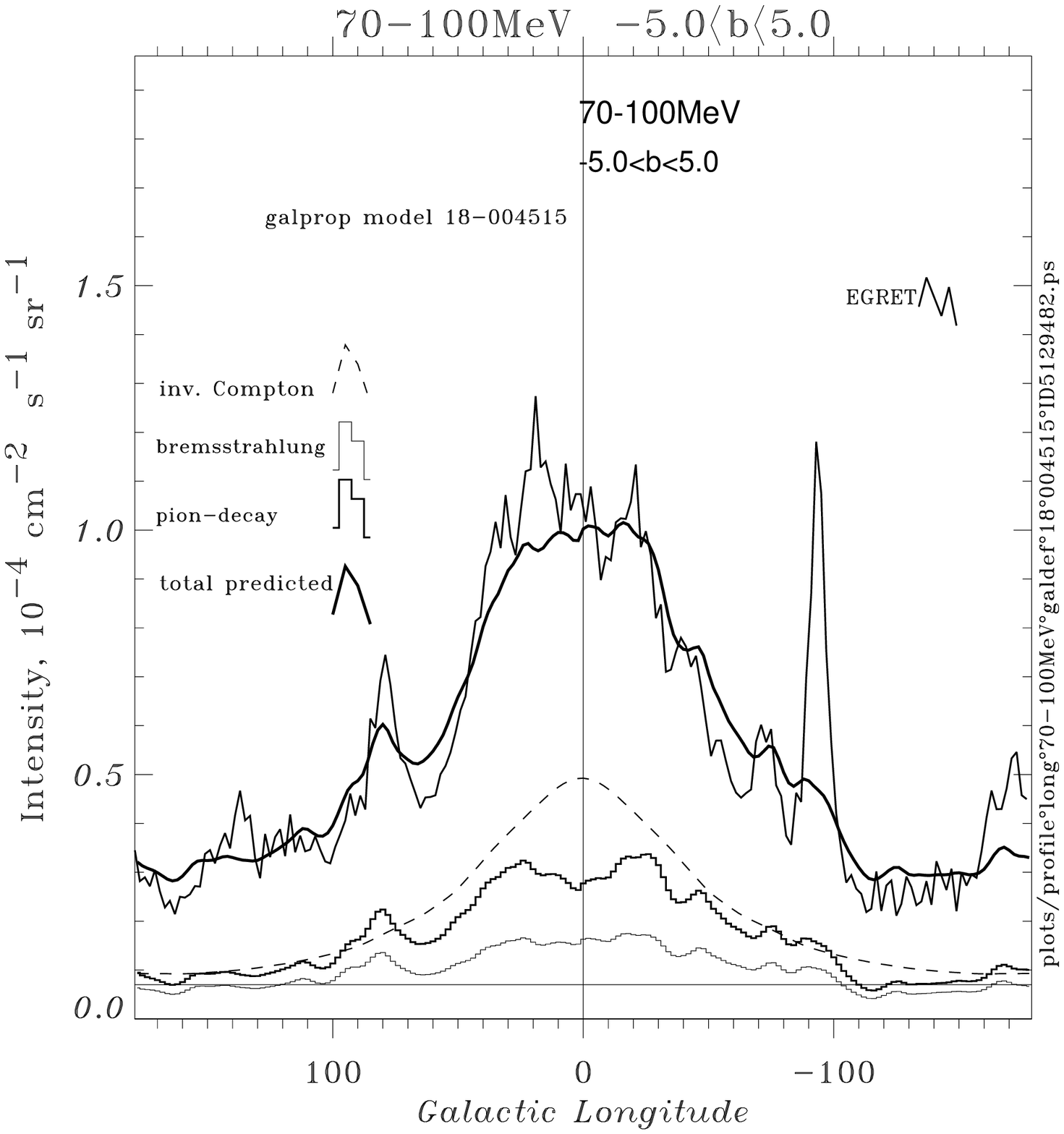,%
         height=\fhb,width=\fwb,clip=}}}
   \end{picture}
\small Fig. 2.
{\it Left:} Latitude distribution for 70--100 MeV as measured by
EGRET, compared to reacceleration model with {\it hard electron} spectrum.
{\it Right:} Longitude distribution for $|b|<5^\circ$.
\end{figure}

\begin{figure}[thb!]
   \begin{picture}(148,64)(5,-61)
      \put(0,0){ \makebox(75,0)[tl]{ \psfig{file=fig3a.ps,%
         height=\fhb,width=\fwb,clip=}}}
      \put(75,0){ \makebox(75,0)[tl]{ \psfig{file=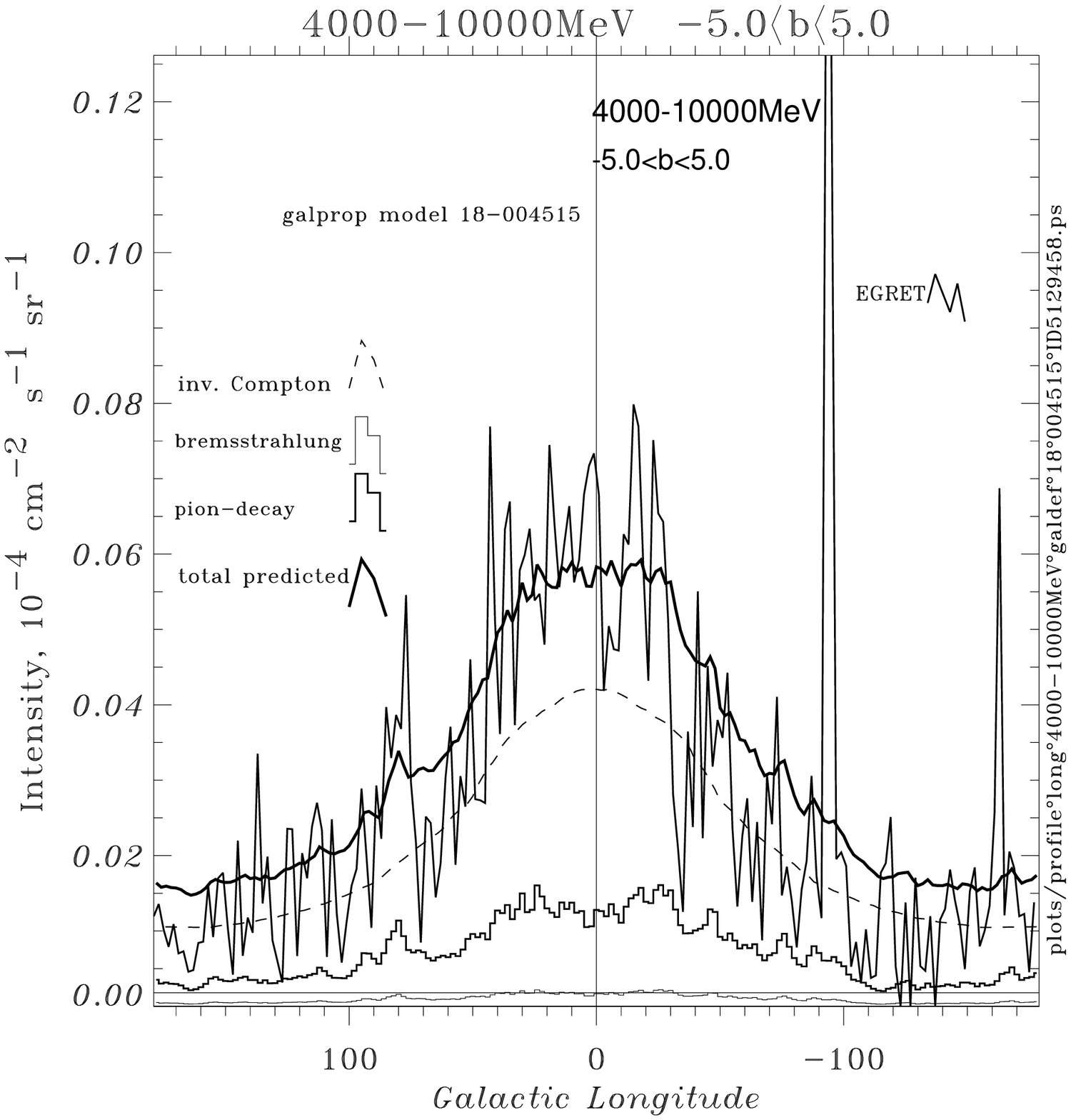,%
         height=\fhb,width=\fwb,clip=}}}
   \end{picture}
\small Fig. 3.
{\it Left:} Latitude distribution for 4000--10000 MeV as measured
by EGRET, compared to reacceleration model with {\it hard electron} spectrum.
{\it Right:} Longitude distribution for $|b|<5^\circ$.
\end{figure}

\begin{figure}[t!]
   \begin{picture}(148,60)(5,-59)
      \put(0,0){ \makebox(75,0)[tl]{ \psfig{file=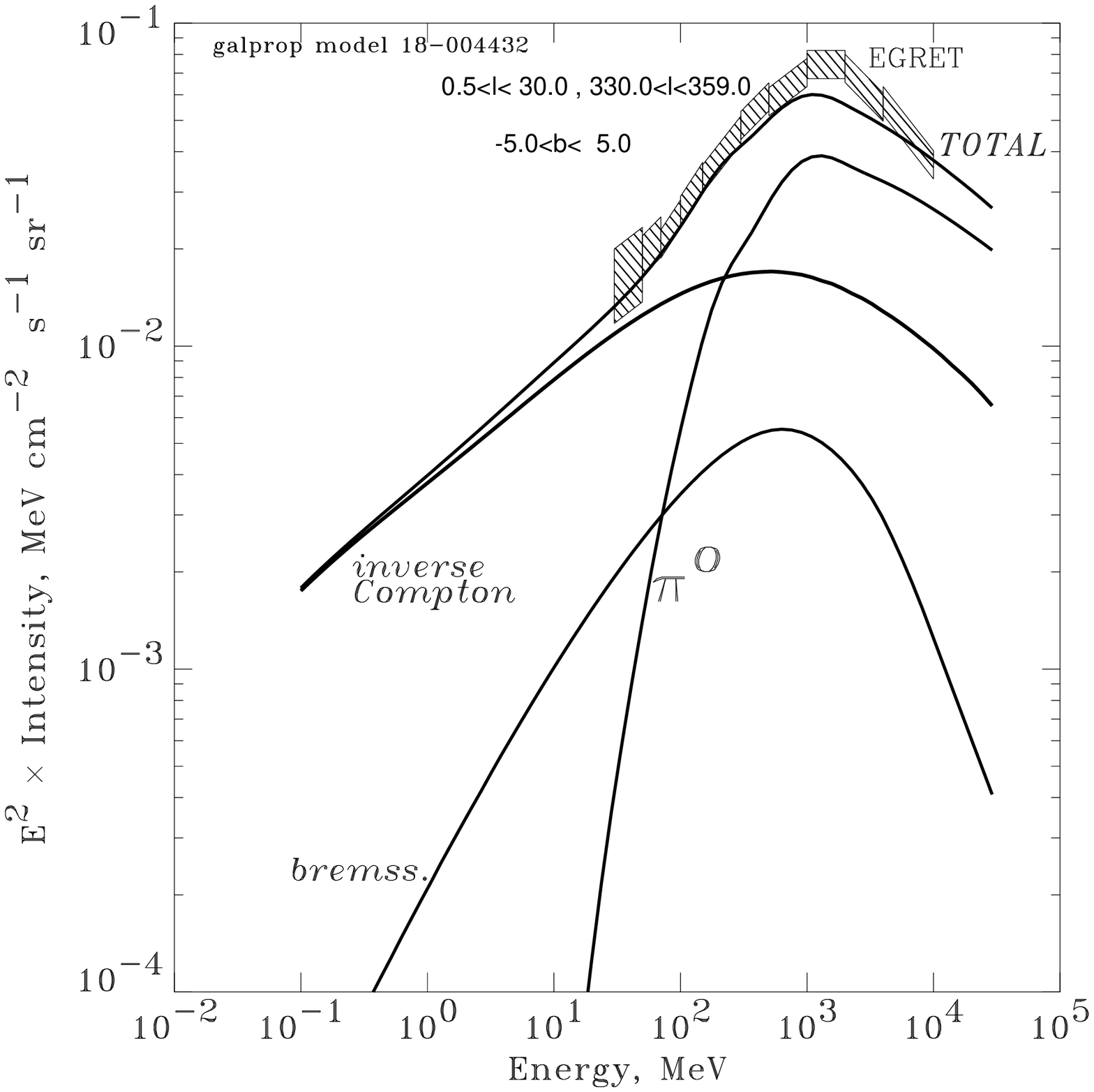,%
         height=\fhb,width=\fwb,clip=}}}
      \put(75,0){ \makebox(75,0)[tl]{ \psfig{file=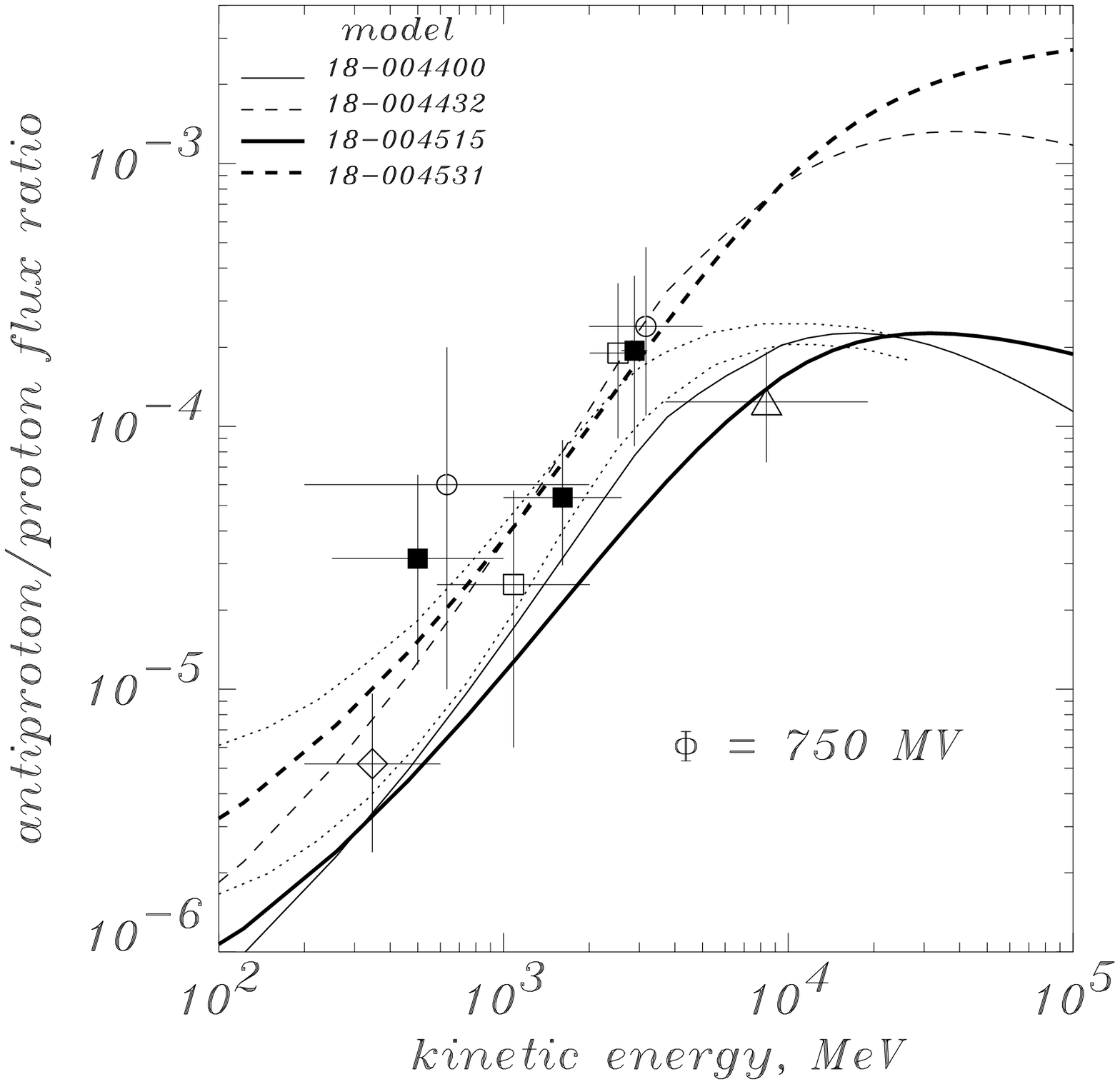,%
         height=\fhb,width=\fwb,clip=}}}
   \end{picture}
\small Fig. 4.
{\it Left:} Gamma-ray spectrum of inner Galaxy as measured by EGRET
(Strong 1996b) compared to model with a {\it hard nucleon} spectrum.
{\it Right:} $\bar{p}/p$ ratio for the `normal' spectrum (solid lines)
and for the hard nucleon spectrum (dashes) used for the \gray
calculation shown on the left. The thick lines show the case with
reacceleration.  Dotted lines: calculations of Simon (1998).  Data
from:  \rule[0pt]{1.5ex}{1.5ex}\ Boezio (1997), {\large $\circ$}
Bogomolov (1987,1990), $\bigtriangleup$ Hof (1996),
\raisebox{0.7ex}{\framebox[1.2ex]{}} Mitchell (1996), $\diamondsuit$
Moiseev (1997).
\end{figure}

\section{Test for a hard nucleon spectrum using antiprotons and positrons.}
Another possible origin for the $>$1 GeV excess could be an
interstellar nucleon spectrum which is harder than observed locally
(e.g., Hunter 1997, Gralewicz 1997, Mori 1997). Figure~4 (left)
illustrates such a possibility; here we have used a nucleon injection
spectrum which is a power law in momentum with index --1.7 (no
reacceleration) giving after propagation a \gray spectrum which agrees
reasonably with the EGRET data.

\begin{wrapfigure}[15]{r}[0mm]{73.2mm}
   \begin{picture}(68,61)(5,-3)
   \put(3,0){\makebox(68,60)[l]%
{\psfig{file=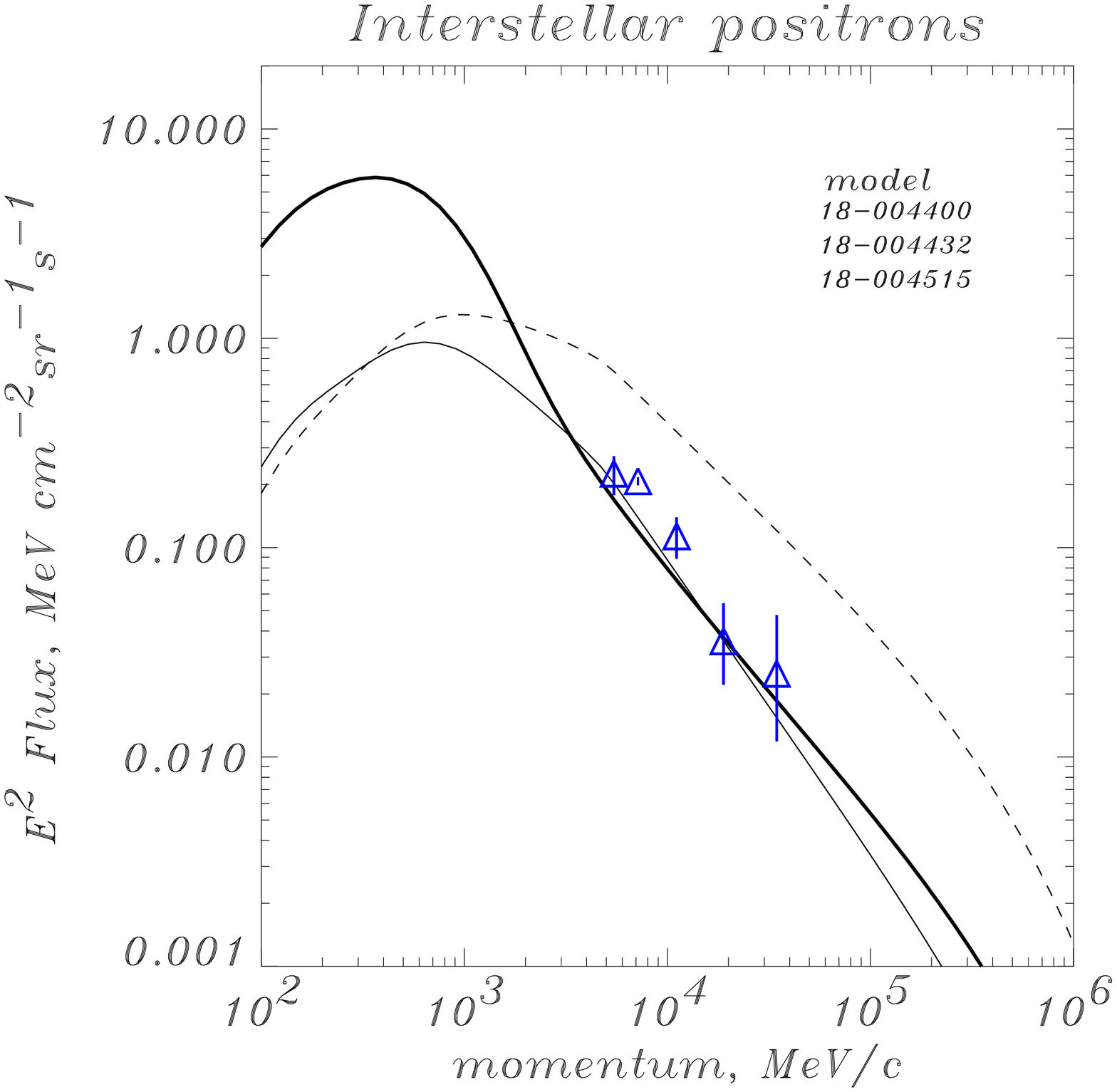,width=\fwc,height=\fhc,clip=}}}
   \end{picture}

\vskip -7mm
\small Fig. 5. Spectra of secondary positrons for `normal' (thin line)
and hard (dashes) nucleon spectra (no reacceleration). 
Thick line: `normal' case with reacceleration.
Data from Barwick (1998).
\end{wrapfigure}

The $\bar{p}/p$ ratio expected for this case and the standard model
compared to recent data is shown in Figure~4 (right).  Our standard
model calculation agrees with that of Simon (1998).  For the case of a
hard nucleon spectrum the ratio is still consistent with the data at
low energies but becomes $\sim$4 times higher at 10 GeV.  Up to 3 GeV
it does not confict with the data with their  large error bars.  It is
however larger than the point at 3.7--19 GeV (Hof 1996) by about
$5\sigma$.  On the basis of the $\bar{p}/p$ data point above 3 GeV we
seem already to be able to exclude the hard nucleon spectrum, but
confirmation of this conclusion must await more accurate data at high
energies.

Figure 5 shows the interstellar positron spectrum, again for the
standard and hard nucleon spectra.  The formalism used is given in
Moskalenko (1998a).  The flux for the standard case agrees with recent
data (Barwick 1998). For the hard nucleon spectrum the flux is higher
than observed by factor $\sim$4; this provides more evidence against a
hard nucleon spectrum.  However this test is less direct than $\bar{p}$
due to the difference in particle type and the large effect of energy
losses.

\smallskip
\secref{References.} 
\setlength{\parindent}{-5mm}
\begin{list}{}{\topsep 0pt \partopsep 0pt \itemsep 0pt \leftmargin 5mm
\parsep 0pt \itemindent -5mm}
\item S.W.~Barwick et al.\ {\it Ap.J.} 498 (1998) 779--789.
\item M.~Boezio et al.\ {\it Ap.J.} 487 (1997) 415--423.
\item E.A.~Bogomolov et al.\ {\it 20th ICRC.} 2 (1987) 72--75.
\item E.A.~Bogomolov et al.\ {\it 21st ICRC.} 3 (1990) 288--290.
\item C.D.~Dermer.\ {\it A\&A.} 157 (1986) 223--229.
\item P.~Gralewicz et al.\ {\it A\&A.} 318 (1997) 925--930.
\item M.~Hof et al.\ {\it Ap.J.} 467 (1996) L33--L36.
\item S.D.~Hunter et al.\ {\it Ap.J.} 481 (1997) 205--240.
\item J.W.~Mitchell et al.\ {\it Phys.Rev.Let.} 76 (1996) 3057--3060.
\item A.~Moiseev et al.\ {\it Ap.J.} 474 (1997) 479--489.
\item M.~Mori.\ {\it Ap.J.} 478 (1997) 225--232.
\item I.V.~Moskalenko and A.W.~Strong.\ {\it Ap.J.} 493 (1998a) 694--707.
\item I.V.~Moskalenko, A.W.~Strong and O.~Reimer.\ {\it A\&A.} (1998b)
      submitted.\ (astro-ph/9808084)
\item M.~Pohl and J.A.~Esposito.\ {\it Ap.J.} 507 (1998) in press.
\item M.~Simon, A.~Molnar and S.~Roesler.\ {\it Ap.J.} 499 (1998) 250--257.
\item A.W.~Strong et al.\ {\it A\&AS.} 120 (1996a) C381--C387.
\item A.W.~Strong and J.R.~Mattox.\ {\it A\&A.} 308 (1996b) L21--L24.
\item A.W.~Strong et al.\ {\it 4th Compton Symp.\ AIP 410.} 
   Ed.\ C.D.~Dermer et al.\ 1198--1202.\ AIP.\ NY.\ (1997).
\item A.W.~Strong and I.V.~Moskalenko.\ {\it 16th ECRS.} (1998) OG-2.5.\
   (astro-ph/9807289)

\end{list}
\end{document}